\documentclass{jfm}
\usepackage{graphicx}
\usepackage{epstopdf, epsfig}
\usepackage{color}
\usepackage{bm}
\usepackage{amsmath}
\usepackage{amssymb}
\usepackage{graphicx}
\DeclareMathOperator{\sech}{sech}
\usepackage{upgreek}
\usepackage{cleveref}
\crefformat{section}{\S#2#1#3}
\crefformat{subsection}{\S#2#1#3}

\shorttitle{Internal wave energy flux from density perturbations in nonlinear stratifications}
\shortauthor{F. M. Lee, M. R. Allshouse, H. L. Swinney and P. J. Morrison}

\title{Internal wave energy flux from density perturbations in nonlinear stratifications}

\author{Frank M. Lee\aff{1},
  Michael R. Allshouse\aff{2,3} \corresp{\email{m.allshouse@northeastern.edu}},
  Harry L. Swinney\aff{2}
 \and Philip J. Morrison\aff{1}}

\affiliation{\aff{1}Institute for Fusion Studies and Department of Physics, University of Texas at Austin, Austin, TX 78712, USA
\aff{2}Center for Nonlinear Dynamics and Department of Physics, University of Texas at Austin, Austin, TX 78712, USA
\aff{3}Department of Mechanical and Industrial Engineering, Northeastern University, Boston, MA, 02115, USA}

\begin{document}
 
\maketitle

\begin{abstract}
Internal gravity wave energy contributes significantly to the energy budget of the oceans, affecting mixing and the thermohaline circulation. Hence it is important to determine the internal wave energy flux  $\bm{J} = p \, \bm{v}$, where $p$ is the pressure perturbation field and $\bm{v}$ is the velocity perturbation field. However, the pressure perturbation field is not directly accessible in laboratory or field observations.  Previously, a Green's function based method was developed to calculate the instantaneous energy flux field from a measured density perturbation field $\rho(x,z,t)$, given a {\it constant} buoyancy frequency $N$.  Here we present methods for computing the instantaneous energy flux $\bm{J}(x,z,t)$ for a spatially varying $N(z)$, as in the oceans where $N(z)$ typically decreases by two orders of magnitude from shallow water to the deep ocean. Analytic methods are presented for computing $\bm{J}(x,z,t)$ from a density perturbation field for $N(z)$ varying linearly with $z$ and for $N^2(z)$ varying as $\tanh (z)$. To generalize this approach to {\it arbitrary} $N(z)$, we present a computational method for obtaining $\bm{J}(x,z,t)$. The results for $\bm{J}(x,z,t)$ for the different cases agree well with results from direct numerical simulations of the Navier-Stokes equations. Our computational method  can be applied to any density perturbation data using the MATLAB graphical user interface EnergyFlux.
\end{abstract}



\section{Introduction}

Ubiquitous internal gravity waves are generated in the oceans by tidal flow over bottom topography and by surface storms~\citep{munk98,alford03,wunsch04}.  The internal waves transmit energy from generation sites to large distances, and ultimately the energy is converted into small-scale mixing. Internal waves are a major contributor of the energy budget of the oceans, and the mechanism for this contribution can be better understood through the energy flux field.  In this paper, we examine the baroclinic energy flux $\bm{J}$ in the presence of a stable background density stratification, 
\begin{align}
\bm{J} = p \, \bm{v}, \label{jpv}
\end{align}
where $p$ is the pressure perturbation from the background hydrostatic pressure field, and $\bm{v}$ is the velocity perturbation from the background velocity flow field. Determining the energy flux requires the simultaneous measurement of both the pressure and velocity perturbation fields.  In numerical investigations of internal waves, these fields are computed~\citep{lamb04, niwa04, zilberman09, king09, king10, gayen10, gayen11, rapaka13}, enabling a direct calculation of the energy flux.  However, In laboratory and field studies simultaneous measurements of the velocity and pressure perturbation fields are usually not possible.

Laboratory observations of internal waves have been made by particle image velocimetry (PIV) \citep{echeverri09, king09, king10, paoletti14}, synthetic schlieren \citep{sutherland99, dalziel00, clark10, allshouse16}, and light attenuation~\citep{dossmann16}. In PIV, neutrally buoyant particles scatter incident laser light, and movies of the scattered light field yield the time-varying velocity field. In the schlieren method, a patterned mask is placed behind a tank that contains the internal waves, and the time-varying distorted image formed by light transmitted through the tank in the transverse direction gives the time-dependent density perturbation field~\citep{sutherland99, dalziel00}.  The light attenuation technique uses dye intensity that varies with depth and moves with the fluid to measure the density field as a function of time~\citep{dossmann16}. These experimental techniques yield the velocity field in the case of PIV and the density perturbation field in the case of synthetic schlieren and light attenuation measurements.  No experimental technique directly yields the pressure perturbation field needed for the energy flux calculation.   

Given the importance of the energy flux and the far-field radiated power obtained by integrating the flux, multiple efforts have been made to calculate the energy flux from experimental measurements.  One method assumed a constant buoyancy frequency and calculated the energy flux (averaged over a tidal period) given only the stream function, thus eliminating the need to measure the pressure perturbation field~\citep{balmforth02}.   The stream function method was subsequently extended to a buoyancy frequency varying exponentially with depth  \citet{lee14}.  Another method applied the polarization relations to density perturbation data to obtain estimates for the velocity and pressure perturbation amplitudes~\citep{clark10}.  This method provided the  energy flux time averaged over a tidal period for a monochromatic plane wave, which is not representative of the complex ocean internal wave field, which has many natural frequencies and spatially varying wave packets.    Finally, an ocean observation analysis technique calculated the pressure perturbation field for measured density profiles, assuming a hydrostatic pressure field; this together with simultaneous velocity measurements allowed the calculation of the energy flux field~\citep{nash05}.  This method assumed that the contribution of dynamic pressure is negligible, which is often not the case in experiments and in high velocity events in the ocean. 

\begin{figure}
  \centerline{\includegraphics[width=.6\textwidth]{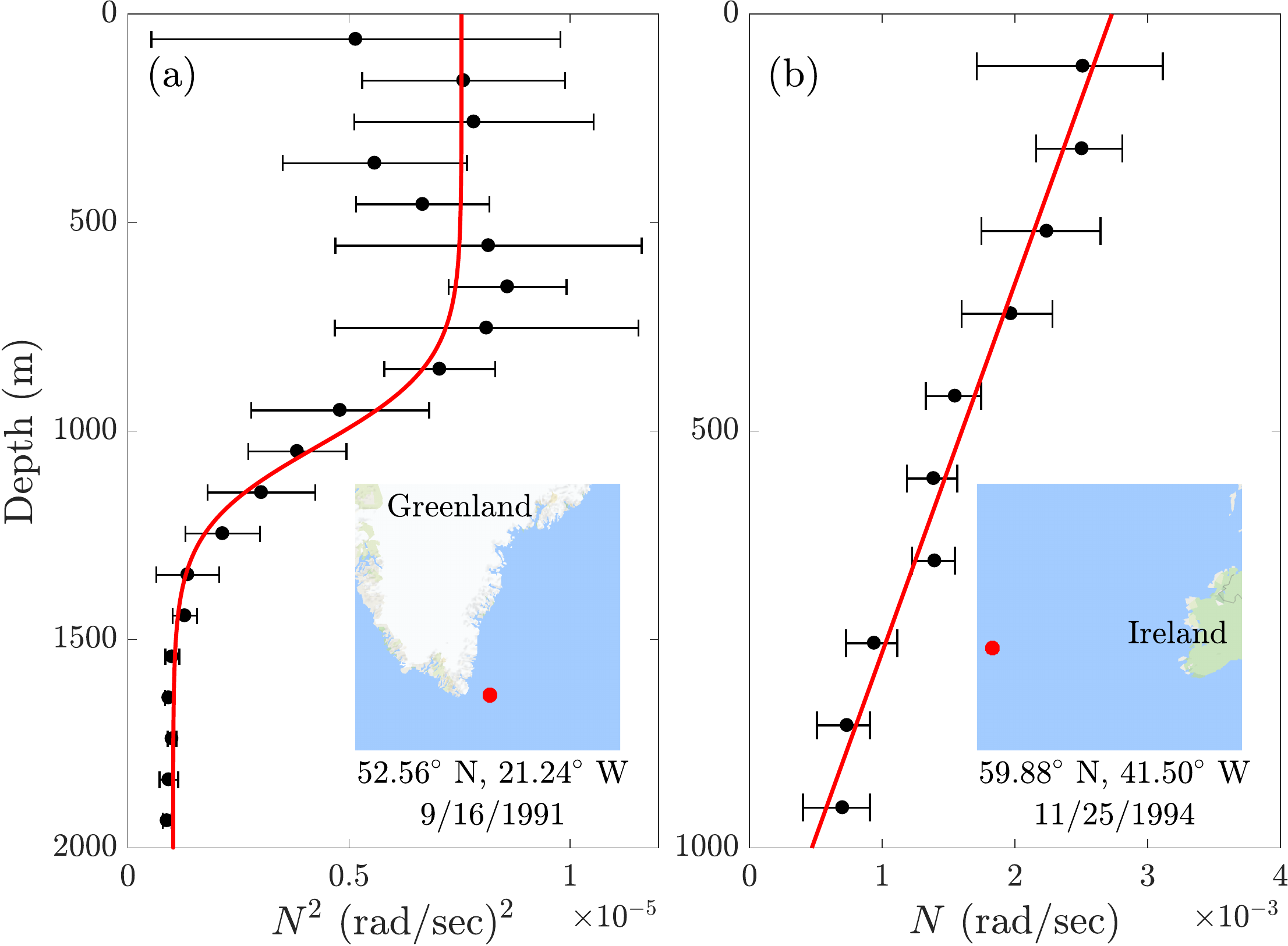}}
  \caption{Two buoyancy frequency profiles from a World Ocean Circulation Experiment data set: (a) A buoyancy frequency squared profile (black) fit to a tanh profile (red).  (b) A buoyancy frequency profile (black) fit to a linear profile (red). The insets show regions 1000 km $\times$ 1000 km that contain the locations (red dots) where the measurements were made. The mean buoyancy frequency (squared for (a)) of bins of stratification measurements is plotted as a function of depth (black dots) with error bars representing two standard deviations from the mean.}
\label{fig:fig1}
\end{figure}

The method presented here for obtaining the instantaneous energy flux from laboratory and field data differs from previous methods that determined the time-averaged energy flux or a global energy conversion rate. Recently a Green's function method was used to calculate  the instantaneous energy flux field from a density perturbation field~\citep{allshouse16} for a fluid with constant buoyancy frequency,
\begin{equation}
N(z) = \sqrt{\frac{-g}{\rho_0}\frac{\partial \rho}{\partial z }},
\label{N}
\end{equation}
but in the oceans $N$ varies significantly with depth, as figure~\ref{fig:fig1} illustrates with data from two locations. 

Recognizing the strength of the Green's function method, we expand that method to accommodate linear $N$ and tanh $N^2$ profiles. These two profiles were selected due to their mathematical properties and their presence in ocean stratifications.  The tanh $N^2$ profile is often a good approximation in the ocean when two near-constant buoyancy-frequency zones are separated by a pycnocline, as figure~\ref{fig:fig1}(a) illustrates.  The linear $N$ profile can occur in the ocean when there is no strong pycnocline, and the buoyancy frequency near the surface decays to a near constant value in the depth, as illustrated in figure~\ref{fig:fig1}(b).

However, as many buoyancy frequency profiles cannot be adequately approximated by either a linear $N$ or a tanh $N^2$ profile as was done in figure \ref{fig:fig1}, the general $N(z)$ case must be treated separately. To account for this, we present a numerical method for computing the instantaneous energy flux field and the integrated far-field internal wave power solely from a density perturbation field, which can have an {\it arbitrary} $N(z)$ profile.

The present work provides a tool for laboratory experiments and field measurements: the calculation of the instantaneous energy flux field from density perturbation data. The method can be applied to ocean density perturbation space-time data when such data becomes available.  The methods presented here and in \citet{allshouse16} provide the {\it instantaneous} rather than time-averaged energy flux field. Thus the resultant energy flux and integrated far-field power include all spectral components, while previous methods provided only the global conversion rates or monochromatic results.

The paper is organized as follows. An outline of our method for obtaining the energy flux from the density perturbation field in a tanh $N^2$ and linear $N$ stratification is presented in \cref{section-theory}.  This method is then verified with numerical simulations in \cref{section-verification}.  A finite difference method for calculating the energy flux for an arbitrary buoyancy frequency profile is presented in \cref{section-arbitrarystratification}, and is applied to an ocean-inspired stratification.  Lastly, conclusions and potential applications of this work are presented in \cref{section-conclusions}.

\section{Theoretical development} 
\label{section-theory}

As this work builds off the theoretical foundation presented in~\citet{allshouse16}, we present the general equations in \cref{section-gen}.  These equations provide the velocity perturbation field as a function of the density perturbation, and a functional relationship between the density and pressure perturbation fields is established.  For an analytic tanh $N^2$ and linear $N$, we calculate the Green's function in \cref{section-tanhprofile} and \cref{section-linearprofile}, respectively.

\subsection{Generalities}
 \label{section-gen}

Our goal of obtaining the energy flux \eqref{jpv} from the density perturbation field alone requires calculating the pressure perturbation and components of the velocity perturbation from the density perturbation field.  The details of  these calculations were  given in \citet{allshouse16} for a uniform $N$ profile, but here we present a condensed version of  the pressure perturbation calculation,  as  needed for  the calculations for the tanh $N^2$ and linear $N$  profiles. 

We begin   with the linearized Euler equations for perturbation about  a hydrostatic background, 
\begin{eqnarray}
&&\frac{\partial u}{\partial t} = -\frac{1}{\rho_0} \frac{\partial p}{\partial x}\,,  
\qquad
\frac{\partial w}{\partial t} = -\frac{1}{\rho_0} \frac{\partial p}{\partial z} - \frac{\rho}{\rho_0} g\,, \label{dudt}    \\
&&
\frac{\partial \rho}{\partial t} = \frac{N^2 \, \rho_0}{g} w\,,  \qquad
\frac{\partial u}{\partial x} + \frac{\partial w}{\partial z} = 0\,,
 \label{div1}
\end{eqnarray}
where  $u$ and $v$ are the horizontal and vertical components of the velocity perturbation, respectively, $p$ is the pressure perturbation, $\rho$ is the density perturbation, $\rho_0$ is the hydrostatic background density profile, and $N$ is the buoyancy frequency.
By manipulating  \eqref{dudt} and \eqref{div1} we obtain an equation for the pressure perturbation in terms of the density perturbation,
\begin{align}
\frac{\partial^2 p}{\partial x^2} + \frac{\partial^2 p}{\partial z^2} + \frac{N^2}{g} \frac{\partial p}{\partial z}
=& - N^2 \rho  -  g \frac{\partial \rho}{\partial z}\,. 
\label{peqn}
\end{align}
First, we solve this equation for $p$ assuming we have the measured $\rho$.

Equation \eqref{peqn} is brought  into a convenient form by first applying the following transformation:
\begin{align}
p(x,z) = q(x,z)\, T(z)
\label{pressure}
\end{align}
where
\begin{align}
T(z) = \exp{ \left[ -\frac{1}{2g} \int^{z}{dz' N^2(z')}\right] }
 \label{Ttrans}
\end{align}
and then Fourier expanding in $x$, yielding
\begin{align}
\frac{d^2 Q}{d z^2} - \left( k^2 + \frac{N}{g} \frac{dN}{dz} + \frac{N^4}{4\,g^2} \right) Q = -F.
\label{Qeq}
\end{align}
Here $F(z;k)$ and $Q(z;k)$  denote the Fourier components of
\begin{align}
f(x,z) = \frac{1}{T(z)} \left( N^2 \rho  +  g \frac{\partial \rho}{\partial z} \right)
\end{align}
and $q(x,z)$, respectively.

We solve \eqref{Qeq} for $Q$ given $F$ by obtaining the Green's function for the Fourier components, which satisfies
\begin{align}
\frac{d^2}{dz^2}G(z,z';k) - \left( k^2 + \frac{N}{g} \frac{dN}{dz} + \frac{N^4}{4g^2} \right) G(z,z';k)   = 0,\quad z \neq z'\,, 
\label{Geqn0}
\end{align}
with a no-flux condition in the $z$ direction at the top and bottom of the domain,  
\begin{align}
\left( \frac{dG}{dz}  - \frac{N^2}{2g} G \right) \bigg|_{z=0,\,h} = 0 \,, 
\label{Gbdy}
\end{align}
and the Green's function matching conditions,
\begin{align}
G^+(z') =& G^-(z') \label{Gmatching1} \\ 
\frac{dG^+}{dz}(z') =& \frac{dG^-}{dz}(z') + 1 \,. \label{Gmatching2}
\end{align}
Thus, given profiles for the buoyancy frequency $N$ and  source term $f$,  the pressure perturbation is given by  the following expression:
\begin{align}
p(x,z) = \mathrm{Re} \, \bigg\{ - \frac{2}{l} \, T(z) \sum_{k} { e^{- ikx} \int_{0}^{h}{ dz' \, G(z,z';k) \, \int_{0}^{l}{dx' \, f(x',z') \, e^{ikx'}} }} \bigg\}, \label{greenspressure}
\end{align}
where $k = 2 \pi n / l$, $l$ is the width of the system, and $n$ is a positive integer.

Next, we obtain the components of the velocity perturbation.  The vertical component follows  by inverting the first equation of \eqref{div1} yielding, 
\begin{align}
w = \frac{g}{N^2 \, \rho_0} \frac{\partial \rho}{\partial t}, \label{weqn}
\end{align}
and the horizontal component is obtained by using the vertical velocity perturbation \eqref{weqn} and the incompressibility condition, the second equation of \eqref{div1}, which gives the differential equation
\begin{align}
\frac{\partial u}{\partial x} = -\frac{\partial}{\partial z} \left( \frac{g}{N^2 \rho_0} \frac{\partial \rho}{\partial t} \right). \label{ueqn}
\end{align}

None of these calculations depend on the particular form of the buoyancy frequency profile, so it is possible to perform all the necessary expressions for calculating the energy flux from $\rho$ alone in a general stratification.  To calculate analytically the Green's function for \eqref{Geqn0}, it is necessary that  the functional form of the buoyancy frequency profile be specified.  \citet{allshouse16} investigates the particular case where the buoyancy frequency is constant resulting in a Green's function that is exponential.  In \cref{section-tanhprofile} we present the calculations for obtaining the pressure perturbation for the tanh $N^2$ profile, and in \cref{section-linearprofile} we present the analogous calculation for the linear $N$ profile.

\subsection{The tanh profile}
 \label{section-tanhprofile}


The buoyancy frequency profile we assume in this section is given by
\begin{align}
N^2(z) =& \frac{N_1^2 + N_2^2}{2} + \frac{N_2^2 - N_1^2}{2} \tanh\big(\alpha (z-z_t)\big)  \\
\equiv& \eta_+ + \eta_- \tanh\big(\alpha (z-z_t)\big), \label{Ntanh}
\end{align}
because this gives a convenient form for $N\,dN/dz$. Here $\alpha$ controls the transition width between the two buoyancy frequency values $N_1$ and $N_2$, and $z_t$ is the midpoint of the transition. Note, for large $\alpha$ \eqref{Ntanh} approximates a  two-layer $N^2$ profile, which we will investigate in \cref{section-tanhverification}. We assume that $N^4/4g^2$ in \eqref{Ntanh} is negligible. For low mode numbers, outside of the transition region, $k^2$ ($\sim 10^{0}-10^{1}$ in MKS) is much larger than $N^4/4g^2$ ($\sim 10^{-2}$), and near the transition region $k^2$ is roughly the same order as $(N/g)(dN/dz)$. For larger mode numbers $k^2$ is the dominating term. Thus for simplicity we keep $k^2$ and $(N/g)(dN/dz)$ and drop $N^4/4g^2$ for all modes.  Upon substituting the analytic stratification into the Green's function equation, \eqref{Geqn0} becomes
\begin{align}
\frac{\partial^2}{\partial z^2}G(z,z';k) - \left( k^2 + \frac{\alpha\, \eta_-}{2 \,g}\sech^2(\alpha (z-z_t)) \right) G(z,z';k)   
= 0,\quad z \neq z' \,.
\label{Geqntanh}
\end{align}
Equation \eqref{Geqntanh} is of the form of a well-studied (e.g.\ \cite{epstein30,poeschl33,lekner07}) time-independent Schr{\"o}dinger equation.

With the  dimensionless coordinate transformation 
\begin{align}
z = z_t + \frac{1}{\alpha} \tanh^{-1}y, \label{zytanh}
\end{align}
equation  \eqref{Geqntanh} becomes 
\begin{align}
(1 - y^2 ) \frac{d^2\bar{G}}{dy^2} -2\,y \frac{d\bar{G}}{dy}  + \left( \nu ( \nu + 1) - \frac{\mu^2}{1 - y^2} \right) \bar{G}   = 0,\quad y \neq y' \,, \label{GLegendre}
\end{align}
where the dimensionless Green's function $\bar{G}$ is given by
\begin{align}
G(z(y)) = \frac{1}{\alpha}\bar{G}(y) 
\end{align}
and   the parameters $\nu$ and $\mu$ are given by
\begin{align}
\nu_{\pm} = -\frac{1}{2} \pm \frac{1}{2} \sqrt{ 1 - \frac{2\,\eta_-}{\alpha \,g} }\,, 
\qquad \mu = \frac{k}{\alpha} \,.
\end{align}
Thus the transformation takes  \eqref{Geqntanh} into the associated Legendre equation \eqref{GLegendre}, which has the 
two linearly independent solutions  $P^{\mu}_{\nu}(y)$ and $Q^{\mu}_{\nu}(y)$, the associated Legendre functions of the first and second kind, respectively. Then, solving \eqref{GLegendre} with the boundary conditions \eqref{Gbdy} and the matching conditions \eqref{Gmatching1} and \eqref{Gmatching2} gives the Green's function,
\begin{align}
\bar{G}(y,y') =\frac{1}{ D \, W }
\begin{cases}
\bigg( \Phi_2 P^{\mu}_{\nu}(y') + \Pi_2 Q^{\mu}_{\nu}(y') \bigg) \bigg( \Phi_1 P^{\mu}_{\nu}(y) + \Pi_1 Q^{\mu}_{\nu}(y) \bigg), & y < y' \\
\bigg( \Phi_1  P^{\mu}_{\nu}(y') + \Pi_1 Q^{\mu}_{\nu}(y') \bigg) \bigg( \Phi_2  P^{\mu}_{\nu}(y) + \Pi_2 Q^{\mu}_{\nu}(y) \bigg), & y > y' \,.
\end{cases}
\label{GreenLegendreFull}
\end{align}
Here
\begin{align}
D = -
\begin{vmatrix}
    \Pi_1 & \Pi_2 \\
    \Phi_1 & \Phi_2
\end{vmatrix}\,, 
\hspace{1cm}
W = 2^{2 \mu} \frac{ \Upgamma( \frac{\nu + \mu + 2}{2} ) \Upgamma( \frac{\nu + \mu + 1}{2} )  }
{ \Upgamma( \frac{\nu - \mu + 2}{2} ) \Upgamma( \frac{\nu - \mu + 1}{2} ) } \,,
\end{align}
\begin{align}
\Pi_{1,2} = \frac{dP^{\mu}_{\nu}}{dy} (y_{0,h}) -  \frac{N_{1,2}^2}{2\,g\,(1-y_{0,h}^2)} P^{\mu}_{\nu}(y_{0,h}) \,, 
\end{align}
\begin{align}
\Phi_{1,2} = - \frac{dQ^{\mu}_{\nu}}{dy} (y_{0,h}) + \frac{N_{1,2}^2}{2\,g\,(1-y_{0,h}^2)} Q^{\mu}_{\nu}(y_{0,h}) \,.
\end{align}
Note, the transformation factor $T(z)$ for  this case is given by 
\begin{align}
T(z) = \left\lbrace \frac{ \cosh[\alpha(z_0-z_t)] }{ \cosh[\alpha ( z-z_t )] } \right\rbrace^{ \eta_{-} / (2 \, \alpha \,g ) } \, \exp{ \left[ \frac{\eta_{+}(z_0-z)}{2\,g} \right]}.
\end{align}


\subsection{The linear profile}
 \label{section-linearprofile}

%
%
%

The calculations for the linear $N$ profile are similar to those of \cref{section-tanhprofile}, so we only highlight the important differences. The linear profile for the buoyancy frequency is given by
\begin{align}
N(z) = \frac{dN}{dz} (z-z_t) \equiv N'(z-z_t)\,, \label{linearNprofile}
\end{align}
where  $z_t$ is now the location where the buoyancy frequency becomes zero. We again neglect $N^4/4g^2$ ($\sim 10^{-2}$) in comparison to $k^2$ and $(N/g)(dN/dz)$ ($\sim n^2$ and $\sim 1$, respectively) as we insert \eqref{linearNprofile} into \eqref{Geqn0}.  For the linear $N$ profile, instead of \eqref{Geqntanh} we obtain
\begin{align}
\frac{\partial^2}{\partial z^2}G(z,z';k) - \left( k^2 + (N')^2 \frac{(z-z_t)}{g} \right) G(z,z';k)   = 0,\quad z \neq z' \,. \label{Geqn_linN}
\end{align}

With the coordinate transformation
\begin{align}
z = z_t - g \, k^2 (N')^{-2} + g^{1/3}(N')^{-2/3}\, y \,,\label{zylinear}
\end{align}
where once again $y$ is a dimensionless coordinate variable, equation  \eqref{Geqn_linN} becomes
\begin{align}
\frac{d^2}{dy^2} \bar{G}(y) - y \, \bar{G}(y) = 0, \quad y \neq y'  \,,\label{Gy}
\end{align}
which is the Airy equation with the  two independent solutions  $Ai(y)$ and $Bi(y)$,  the Airy functions of the first and second kind, respectively. Then,  the dimensionless Green's function is given by
\begin{align}
\bar{G}(y,y') =\frac{\pi}{D}
\begin{cases}
\bigg( \beta_2 Ai(y') + \alpha_2 Bi(y') \bigg) \bigg( \beta_1 Ai(y) + \alpha_1 Bi(y) \bigg), & y < y' \\
\bigg(  \beta_1  Ai(y') + \alpha_1 Bi(y') \bigg) \bigg( \beta_2  Ai(y) + \alpha_2 Bi(y) \bigg), & y > y' \,,
\end{cases}
\label{GreenAiryFull}
\end{align}
which when given dimensions becomes
\begin{align}
G(z(y)) = g^{1/3}(N')^{-2/3} \bar{G}(y).
\end{align}
Here 
\begin{align}
D = -
\begin{vmatrix}
    \alpha_1 & \alpha_2 \\
    \beta_1 & \beta_2
\end{vmatrix}\,,
\end{align}
\begin{align}
\alpha_{1,2} = \frac{dAi}{dy}(y_{0,h}) - \frac{1}{2}\, g^{-2/3} (N')^{4/3}(z_{0,h} - z_t)^2 Ai(y_{0,h}) \,,
\end{align}
\begin{align}
\beta_{1,2} = -\frac{dBi}{dy}(y_{0,h}) + \frac{1}{2}\, g^{-2/3} (N')^{4/3}(z_{0,h} - z_t)^2 Bi(y_{0,h}) \,, 
\end{align}
where $z_0$ and $z_h$ are the coordinates of the bottom and top of the domain, respectively, and $y_0$ and $y_h$ are the corresponding transformed coordinates. The transformation factor $T(z)$ in this case is given by
\begin{align}
T(z) = \exp\left\lbrace \frac{(N')^3}{6\,g} \, \left[ ( z_0-z_t )^3 - ( z-z_t )^3  \right] \right\rbrace.
\end{align}

\section{Analysis verification} \label{section-verification}

To verify the Green's function analysis in \cref{section-theory}, we compare those predictions with results for the energy flux obtained from direct numerical simulations of the Navier-Stokes equations.  The simulations are described in \cref{section-simulation}.  The simulated velocity perturbation, pressure perturbation, and energy flux fields of internal waves in a stratified fluid are compared with the predictions from the analyses for a tanh $N^2$ profile in \cref{section-tanhverification} and for a linear $N$ profile in \cref{section-linearverification}.

\subsection{Simulation of the density perturbation field} \label{section-simulation}

To verify the Green's function method, we perform direct numerical simulations of the Navier-Stokes equations in the Boussinesq approximation.  These simulations provide the density perturbation field needed to calculate the velocity perturbation, pressure perturbation, and energy flux fields. The simulations use the CDP-2.4 algorithm, which is a finite volume solver that implements a fractional-step time-marching scheme~\citep{ham04, mahesh2004}.  This code has previously been used to simulate internal waves and has been validated with experiments~\citep{king09,lee14,dettner13,zhang14,paoletti14,allshouse16}.

\begin{figure}
  \centerline{\includegraphics[width=.7\textwidth]{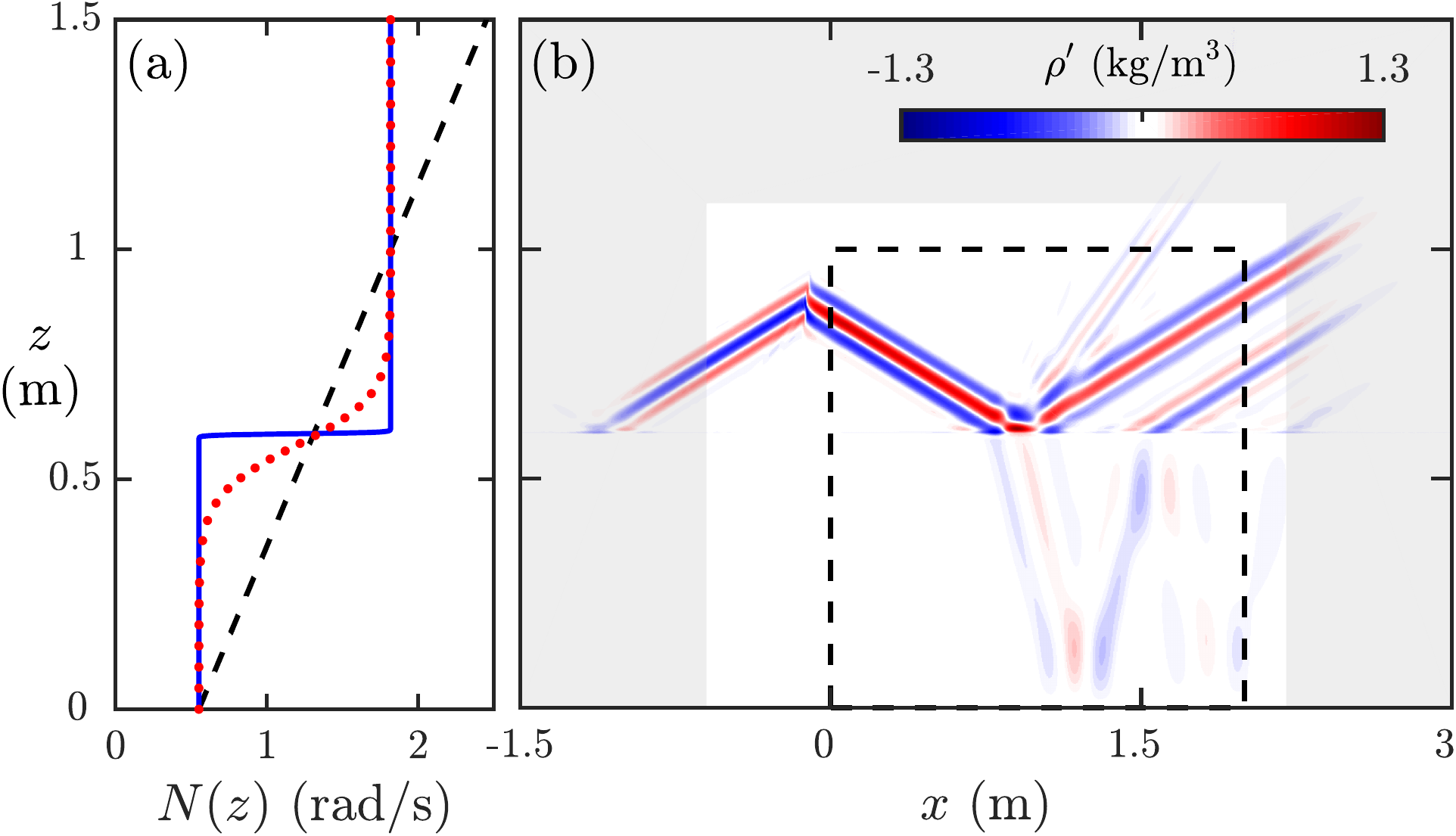}}
  \caption{(a) The $N$ profile of a broad-transition tanh $N^2$ (dotted red curve) and a narrow-transition tanh $N^2$ profile (solid blue curve). The dashed black line is a linear $N$ profile.  (b) The simulation domain and density perturbation field for the narrow transition tanh $N^2$ internal wave field.  Rayleigh damping is applied in the gray region of the field.  The sub domain used for analysis is bound with a black dashed line.}
\label{fig:fig2}
\end{figure}

Our two-dimensional simulations span the domain $x\in~[-1.5, 3]$~m and $z\in~[0, 1.5]$~m.  The simulation solves for the total density $\rho_T$, pressure $p_T$, and velocity $\bm{u}_T$:
\begin{align}
\frac{\partial \bm{u}_T}{\partial t} + \bm{u}_T\cdot\nabla\bm{u}_T &= -\frac{1}{\rho_{00}}\nabla p_T + \nu_w \nabla^2 \bm{u}_T - \frac{g \rho_T}{\rho_{{00}}}\bm{\hat{z}}, \\
\frac{\partial \rho_T}{\partial t} + \bm{u}_T\cdot\nabla\rho_T &= \kappa_s \nabla^2 \rho_T, \nabla \cdot \bm{u}_T = 0,
\end{align}
where $\rho_{00}=1000 $ kg/m$^3$ (density of water), $\nu_w=10^{-6}$ m$^2$/s (kinematic viscosity of water at $20^o$C), and $\kappa_s=2\times10^{-9}$ m$^2$/s (the diffusivity of NaCl in water).  The system is initially at rest and the prescribed density field is unperturbed.  The initial density field is analytically derived from  the buoyancy frequency profiles presented in figure 2(a).  The boundary conditions at the bottom and top are no slip and free slip, respectively.  The left and right boundaries are set to be periodic; however, Rayleigh damping is used along the perimeter of the domain (gray region in figure 2(b)), thus forcing the velocity to be negligible at the left and right boundary.

The internal wave beam is produced by using a momentum source that forms a rectangle with height $0.15$ m and width $0.04$ m, centered at $(-0.02, 0.8)$ m and rotated to match the internal wave beam angle corresponding to the buoyancy frequency at $z=0.8$ m.  The wave beam velocity imposed is 
\begin{align}
\bm{u}_{T} &= \omega A(z') \sin(\omega t - k_z z')\bm{\hat{x'}} \label{uexp},
\end{align}
with an amplitude profile given by
\begin{align}
A(z') &= \exp(-(z')^2/0.0022), \label{Az}
\end{align}
where the lengths are in meters, the rotated coordinates $x'$ and $z'$ correspond to the beam tangent and normal coordinates centered at $\bm{x}=(-0.02,0.8)$ m, respectively, $\omega=2\pi/13$ rad/sec and $k_z=8245$ m$^{-1}$.  A time step size $\delta t=0.0025$~s (5200 steps per period) is sufficient for temporal convergence. Spatial convergence is obtained using an unstructured mesh with resolution $\delta x\approx 0.0014$ m inside the region $x\in [-0.8, 1.80]$ m, $y \in [0.5, 1.1]$ m.  This high resolution region contains the beam generation, the density gradient transition for the tanh $N^2$ profiles, and generation of any additional beams.  The resolution outside of this region grows to $\delta x\approx 0.0025$ m near the boundaries.  Changes in the velocity field are less than $1\%$ when spatial and temporal resolutions are doubled.

The density perturbation field for the case where we have a rapid change in buoyancy frequency (blue line in figure 2(a)) is presented in figure 2(b).  The internal wave beam is generated at $(-0.02,0.8)$ m and produces a beam propagating to the right that is the focus of our studies and a beam propagating to the left which is damped out by the Rayleigh damping.  The beam propagating down to the right reaches the interface at $z=0.6$ m at which point three beams are produced: a reflected beam to the top right at the same angle to the horizontal as the incoming beam, a transmitted beam to the bottom right that has a different angle, and a reflected second harmonic beam at approximately twice the incoming angle.  This particular snapshot is shown after 23.06 periods of forcing, which is sufficient for the beam in the region of interest to reach steady state.

\begin{figure}
  \centerline{\includegraphics[width=\textwidth]{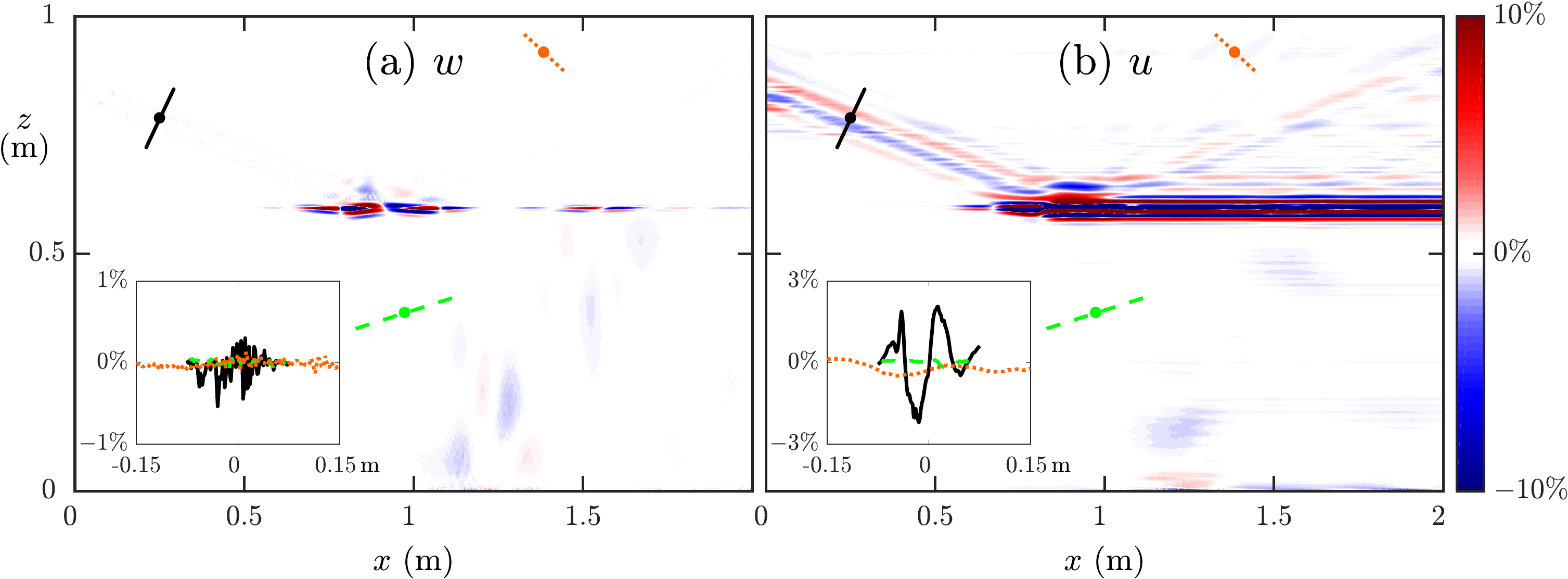}}
  \caption{The beam-normalized percent difference between the Green's function method and the simulations for (a) $w$ and (b) $u$. The insets show the percent difference across the beam for three transects: the incoming beam (solid black), the transmitted  beam (dashed green), and the second harmonic beam (dotted orange).}
\label{fig:fig3}
\end{figure}

\subsection{Tanh $N^2$ profile analysis verification} \label{section-tanhverification}

The vertical \eqref{div1}a and horizontal \eqref{ueqn} components of the velocity and the pressure perturbation calculated from the density perturbation using the Green's function \eqref{GreenLegendreFull} for the tanh $N^2$ profile are verified by comparison with the direct numerical simulations described in \cref{section-simulation}. For large $\alpha$ the tanh $N^2$ profile can be approximated as a two-layer $N$ system, as illustrated in figure 2(a), where $\alpha = 4$, corresponding to a transition thickness of 0.01 m for a $95\%$ change in $N^2$; this is at least an order of magnitude smaller than the the beam width and domain height. Henceforth the large $\alpha$ case is called the ``narrow-transition'' tanh $N^2$ profile.

The difference between the density-perturbation-based velocity perturbation and the simulated velocity perturbation for the narrow-transition tanh $N^2$ profile are presented in figure~\ref{fig:fig3}; in most of the domain, the difference is less than 3\% (with respect to the beam amplitude), except in the transition region at $z = 0.6$ m where there is a significant amount of nonlinearity.  Because the horizontal velocity perturbation is found by solving an ODE on constant $z$ levels \eqref{ueqn}, the patch of large vertical velocity perturbation error is propagated horizontally from the reflection site; thus the region of error is larger for $u$. Despite this nonlinearity, the error is small in most of the domain. 

\begin{figure}
  \centerline{\includegraphics[width=\textwidth]{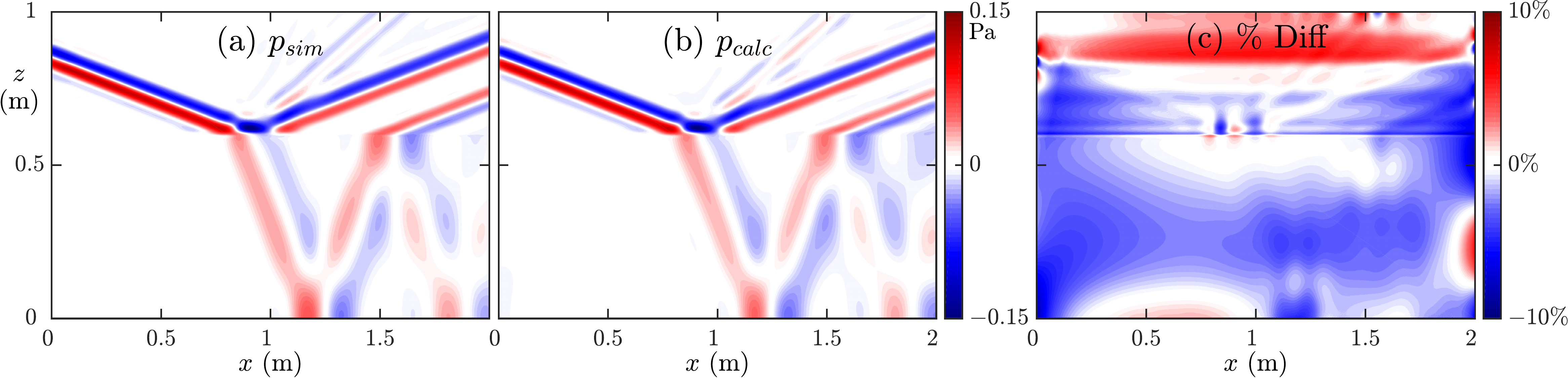}}
  \caption{The pressure perturbation field from (a) the direct numerical simulation and (b) the Green's function calculation from the density perturbation.  (c) The beam-normalized percent difference between the two methods.}
\label{fig:fig4}
\end{figure}

Next we investigate how well the Green's function method calculates the pressure perturbation field from the density perturbation field.  Figure~\ref{fig:fig4}(a) shows the simulated pressure perturbation field, and figure~\ref{fig:fig4}(b) shows the pressure perturbation calculated using the Green's function method. Despite the nonlinearities in the narrow transition layer, the Green's function method, which is based on the linear equations, yields accurate estimates of the pressure perturbation for the reflected, transmitted, and second harmonic beams, as figure~\ref{fig:fig4}(b) illustrates. The beam-normalized percent difference between the calculated and simulated pressure perturbation fields is presented in figure~\ref{fig:fig4}(c).  The calculation is accurate to within $5\%$ over most of the domain, and to better than $10\%$ everywhere except within 0.02 m of where the beam enters the domain. Near the top of the domain, the Green's function method overestimates the pressure perturbation by $4-6\%$, which causes some distortion in the second harmonic, as can be seen around (1.25,0.8) m in figures~\ref{fig:fig4}(a) and \ref{fig:fig4}(b).  The Green's function method underestimates the pressure perturbation in the center of the domain, but the error is less than $5\%$. 

\begin{figure}
  \centerline{\includegraphics[width=\textwidth]{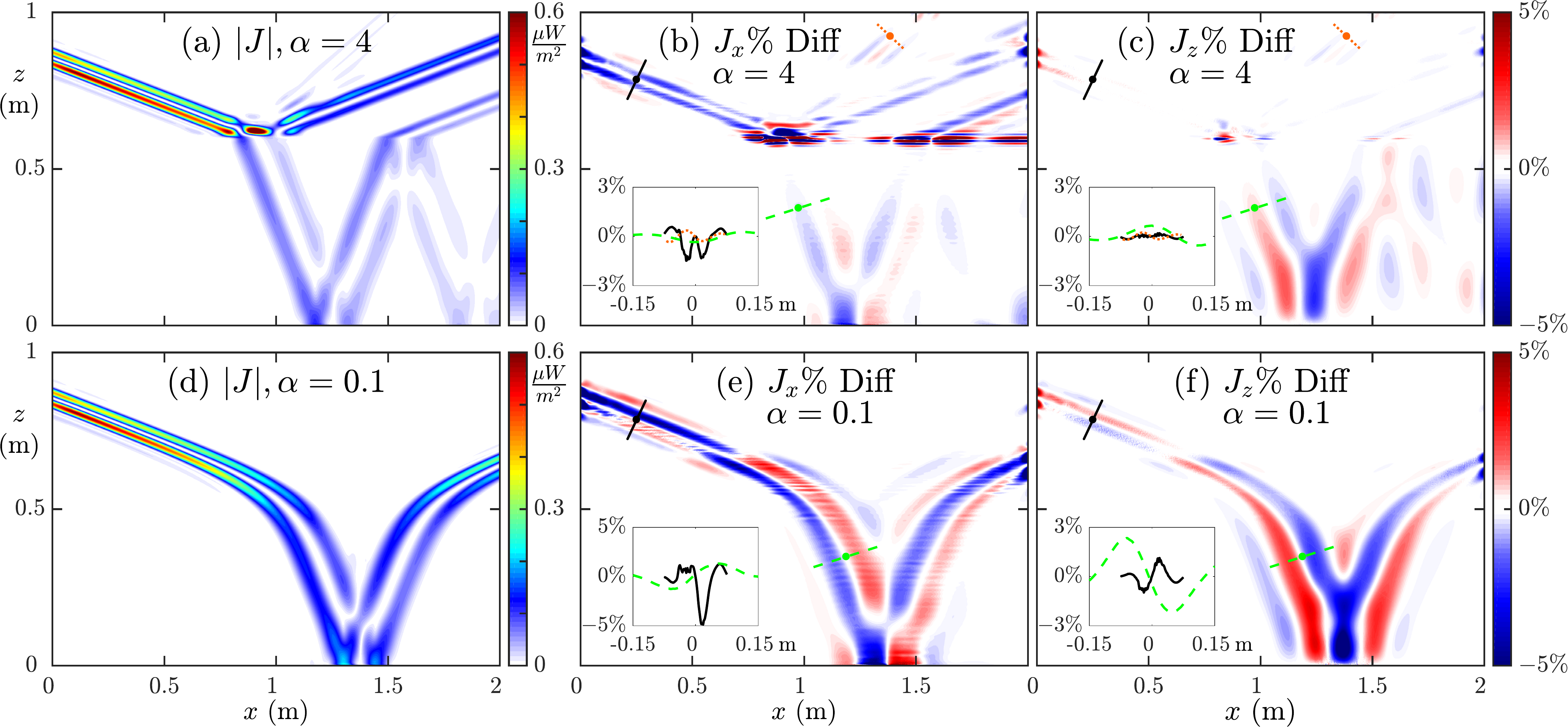}}
  \caption{The energy flux magnitude computed in direct numerical simulations for (a) narrow ($\alpha=4$) and (d) broad ($\alpha=0.1$) tanh $N^2$ transition regions. The beam-normalized percent difference between the $x$-component of the energy flux from the simulations and from the Green's function method is shown in (b) and (e), respectively, for the narrow and broad transition regions, and corresponding results for the $z$-component of the energy flux are in (c) and (f).  For each case an inset shows the difference between the simulations and Green's function methods is less than 3\% for the most of the domain; the insets in each panel show the difference along two or three beam transects.}
\label{fig:fig5}
\end{figure}

Finally, we use the calculated velocity and pressure perturbation fields to compare the energy flux $\bm{J}$ directly from the numerical simulations with the flux computed from the Green's function analysis.  The magnitude of the energy flux from the simulations is presented for the narrow tanh $N^2$ transition regions in figures~\ref{fig:fig5}(a).  For the narrow transition region case the energy flux for the reflected beam is higher than for the transmitted beam and an order of magnitude greater than in the second harmonic.  The beam-normalized percent difference of the horizontal and vertical energy flux are presented in figures~\ref{fig:fig5}(b) and (c), respectively.  Outside of the immediate vicinity of the interface region at $z=0.65$ m the percent difference is less than 3\%.  The accumulated error from multiplying the calculated velocity and pressure perturbation to obtain the flux components is large at the narrow transition interface as a consequence of error in the horizontal velocity perturbation, which is compensated to some extent by a more accurate pressure perturbation calculation at the interface; the error is smaller for $J_z(x,z)$ than for $J_x(x,z)$.  In the lower half of the domain the magnitude of the energy flux is underestimated due to underestimation of the pressure perturbation. The insets show that the error along three beam transects is mostly smaller than $3\%$ for the narrow transition simulation.

We also simulate a tanh $N^2$ profile with a broader transition thickness layer of 0.31 m.  We omit the comparison of the velocity and pressure perturbations for brevity and instead examine the energy fluxes, as shown in figure~\ref{fig:fig5}.  The energy flux field for the broader tanh $N^2$ profile is presented in figure~\ref{fig:fig5}(d).  The internal wave beam passes through the broad transition without reflection because there are no rapid changes in buoyancy frequency.  This smooth transition reduces the nonlinearities so there are significantly smaller errors in the velocity perturbation field and thus the energy flux field as compared to the narrow transition tanh $N^2$.  The magnitude of the energy flux decreases as the beam widens in the bottom of the domain and then increases again as the beam narrows after reflection.  Beam-normalized percent differences are presented for the horizontal and vertical energy flux in figures~\ref{fig:fig5}(e) and (f), respectively.  There is a change of overestimating the energy flux in the top of the domain to underestimating the energy flux in the bottom of the domain.  This is most clearly seen at (0.7,0.7) m where the bands of constant phase change from red to blue and vice versa.  This change is due to the pressure perturbation again being over estimated in the top of the domain and underestimated in the bottom of the domain.  The two insets show that within the beam the percent difference is consistently less than 5\%.  

Figures~\ref{fig:fig5}(a)-(c) demonstrate that our method can handle rapid changes $N^2$, while the broad $N^2$ transition thickness in figures~\ref{fig:fig5}(d)-(f) is more representative of ocean stratifications.  Figure~\ref{fig:fig5}(d) shows the energy flux amplitudes and reveals that the broadening of the transition layer eliminates the reflected and second harmonic beams. Further, the error in the broad transition region in figures~\ref{fig:fig5}(e)-(f) is much smaller than in the narrow transition region figures~\ref{fig:fig5}(b)-(c). The errors are less than $5\%$ except in the narrow transition region.

\subsection{Linear profile analysis verification} \label{section-linearverification}

\begin{figure}
  \centerline{\includegraphics[width=\textwidth]{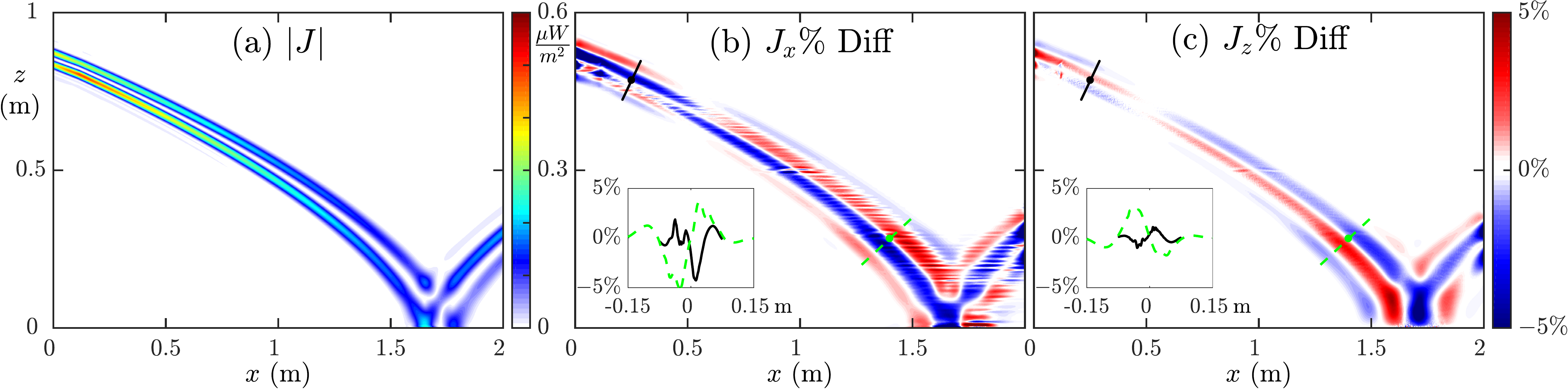}}
  \caption{(a) The energy flux magnitude from the numerical simulation for the linear $N$ profile. The beam-normalized percent difference between simulation and the Green's function method for the (b) $x$ and (c) $z$ components of the energy flux; the difference is less than 3\% for most of the beam, as illustrated by insets showing the difference for two beam transects.}
\label{fig:fig6}
\end{figure}

To verify that the theory for the linear $N$ profile of \cref{section-linearprofile} is valid, we perform simulations analogous to those in \cref{section-tanhverification}.  The energy flux field  in figure~\ref{fig:fig6}(a) demonstrates that the internal wave beam bends more gradually for the linear $N$ profile (figure~\ref{fig:fig6}(a)) as compared to the tanh $N^2$ profiles discussed in \cref{section-tanhverification} (figure~\ref{fig:fig5}(a) and (d)).  This slower change is due to the smaller gradient of the buoyancy frequency for the linear $N$ profile.  Again, because there are no rapid changes in $N$ there are no reflection depths other than the bottom of the system, so nonlinearities are limited to the reflection point at (1.6, 0) m.

We present only the errors in the energy flux calculation; the errors in the velocity and pressure perturbation calculations are qualitatively the same as the results in figures~\ref{fig:fig3} and \ref{fig:fig4}.  Figures \ref{fig:fig6}(b) and (c) show the percent difference of the horizontal and vertical components of the energy flux, respectively. As with the tanh $N^2$ profile comparisons, the errors in the energy flux field are confined to the internal wave beam.  Throughout the beam the difference between the simulation and Green's function method is less than $5\%$, as illustrated by the beam transects in the insets of figures \ref{fig:fig6}(b) and (c); the largest errors occur where the beam enters and leaves the domain and where it reflects off the bottom boundary.  The transition from pressure perturbation overestimation to underestimation is highlighted by the change from red to blue and vice versa near (0.5, 0.75) m.  

\section{Arbitrary stratification analysis} \label{section-arbitrarystratification}

Implementation of the Green's function method works for systems with particular stratifications only when an analytic representation of the Green's function exists.  While some stratifications in the ocean and laboratory may approximately fit to these particular stratification profiles as we show in figure~\ref{fig:fig1}, making this density-perturbation-based calculation more general is necessary for most applications.  To accomplish this generalization, we use a finite difference method to determine the pressure perturbation field.  We present the method in detail along with a comparison between the Green's function method and the finite difference method in \cref{section-finitedifference}.  Then, we apply the finite difference method to an ocean-inspired stratification in \cref{section-finiteverification}.

\subsection{Finite difference method} \label{section-finitedifference}

Since the velocity perturbation calculation does not depend on having an analytic stratification, only the calculation of the pressure perturbation field requires modification for application to general stratifications.  This is accomplished by implementing a numerical solver of the second order differential equation \eqref{Qeq}.   The boundary conditions for this differential equation are analogous to \eqref{Gbdy}:
\begin{align}
\left( \frac{dQ}{dz}  - \frac{N^2}{2g} Q \right) \bigg|_{z=0,\,h} = 0 \,. \label{Qbdy}
\end{align}
We solve equation \eqref{Qeq} using a second-order finite difference method.  The Robin boundary conditions are calculated to second order by adding ghost points to the top and bottom of the domain.  This numerical method is applied to both the real and imaginary  components for every Fourier mode.  After the calculation of $Q(z;k)$ using the finite difference method, the dependence in the $x$-direction is accounted for by multiplying by the particular Fourier mode just as it is done for the Green's function method.  Finally, the transformation \eqref{Ttrans} is performed to determine the contribution to the pressure perturbation field by that particular mode.

\begin{figure}
  \centerline{\includegraphics[width=\textwidth]{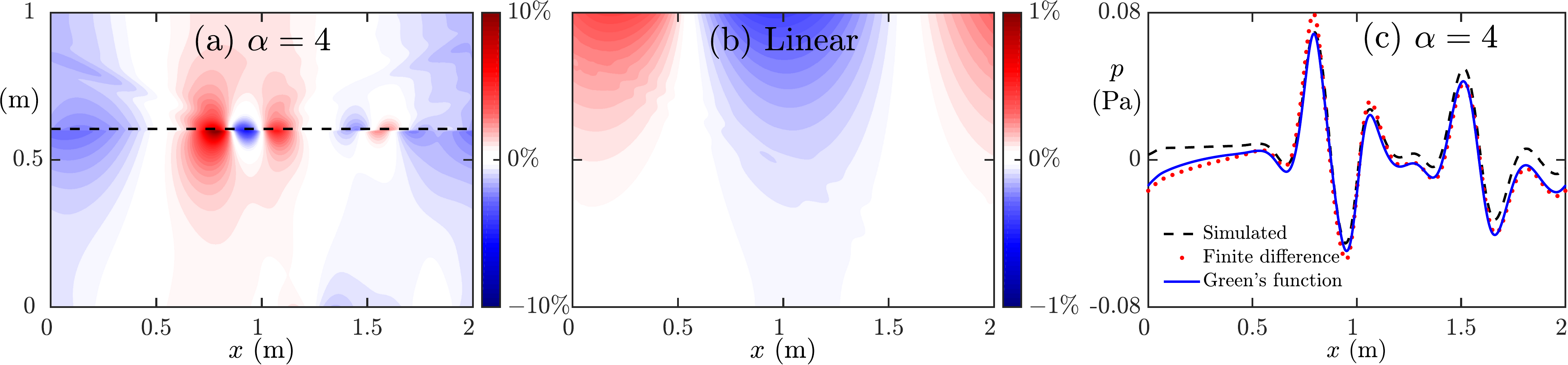}}
  \caption{The percent difference between the Green's function and finite difference pressure perturbation fields for (a) the narrow-transition tanh profile and (b) the linear profile.  (c) Pressure perturbation profiles from the narrow tanh $N^2$ simulation (black), Green's function method (red), and finite difference method (blue) at $z=0.605$ m. }
\label{fig:fig7}
\end{figure}

Applying this strategy to the previous analytic stratifications provides a baseline for comparison to the Green's function method.  The percent difference of the pressure perturbation fields relative to the Green's function results are presented in figure~\ref{fig:fig7}.  This figure shows that the pressure perturbation fields calculated using the finite difference method are everywhere less than 5$\%$ different for the tanh profile and less than $1\%$ different for the linear buoyancy frequency profile when compared to the Green's function pressure perturbation.  The only major discrepancy between the two methods is near the narrow transition in the tanh profile shown in figure~\ref{fig:fig7}(a).  In this region, the Green's function method is consistently more accurate than the finite difference method.  This is highlighted in figure~\ref{fig:fig7}(c) by comparing pressure perturbation profiles just above the transition layer.   The discrepancy is likely due to the Green's function's accurately accounting for the rapid change in the buoyancy frequency when it modifies the coefficients in the calculation of the Legendre functions. The length scale of the transformed coordinate variable $y$ \eqref{zytanh}, is set by the steepness coefficient $\alpha$.  This increases the spacial resolution at rapid transitions.  The finite difference method can only account for variations on the scale of the original data set step size, which, in the case of the narrow transition, is too coarse.

\subsection{Verification of the finite difference method} \label{section-finiteverification}

\begin{figure}
  \centerline{\includegraphics[width=\textwidth]{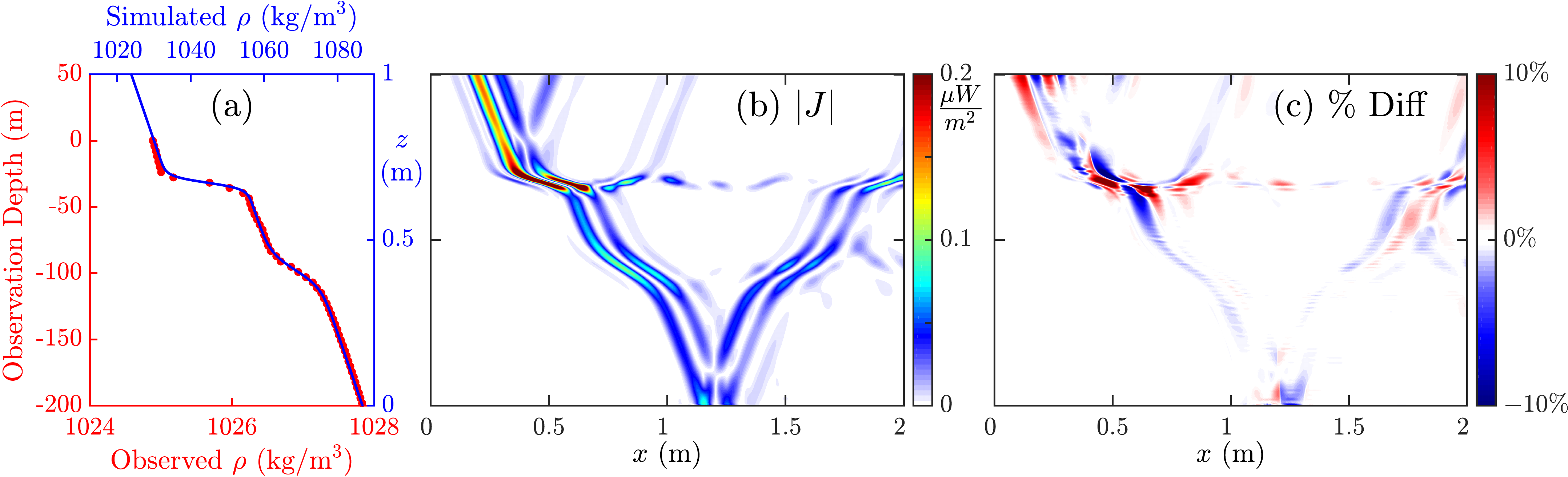}}
  \caption{(a) Density profile from the ocean (red) and the scaled version used for simulation (blue).  (b) The simulated absolute energy flux field.  (c) The beam-normalized percent difference between the simulated and finite difference energy flux fields.}
\label{fig:fig8}
\end{figure}

To further validate the finite difference method, we apply the method to a stratification that does not have a simple analytic function as has been the case in \cref{section-theory} and \cref{section-verification}.  The stratification we simulate is based on a density profile measured in the ocean during the World Ocean Circulation Experiment (WOCE).  The particular profile presented in figure~\ref{fig:fig8}(a) was measured at 165$^{\circ}$ W, 51.5$^{\circ}$ N on September 20$^{th}$, 1994.  This profile features two layers of large density gradient similar to the transitions of the tanh $N^2$ profiles.  The first, more abrupt layer is centered at 30 m below the surface and the second layer is centered at 100 m.  The full profile extends to a depth of 1000 m, but there is little variation in the buoyancy frequency below 200 m.  

In order to simulate the beams in a similar domain and time scale as the analytic stratifications, we rescale the vertical coordinate and the density.  We note that this is done to mimic the actual ocean profile and use it as an inspiration, rather than to model it accurately.  Because the length scale of the transition layers in the simulation are small compared to the length scale of the beams the rescaled simulation done here stress-tests the method.

The first adjustment we make is to provide additional vertical space above the stratification features so that the internal wave beam is fully developed and the resulting reflection off the top of the first transition is visible.  The vertical coordinate is scaled from 200 m to 0.8 m in the simulation.  The density is also modified to increase the buoyancy frequency, so that the values of the buoyancy frequency are comparable to those used in the Green's function verification.  The minimum buoyancy frequency of the scaled density profile is $N=0.55$ rad/s and the maximum value is $N=2.40$ rad/s.  Finally, we shift the location of forcing to be at (0.2,1.2) m to have the internal wave beam enter from the top to demonstrate the flexibility of the domain of measurement.  The time scale and forcing periodicity match the previous simulations.

The magnitude of the energy flux field is presented in figure~\ref{fig:fig8}(b).  There are a number of reflections and transmissions due to the more complicated density profile.  For the first transition layer, the internal wave beam produces reflections off the top and bottom of the pycnocline layer, which can be seen at (0.5, 0.8) m and (1.0,0.8) m, respectively.  In addition to the reflected energy, some of the internal wave energy is trapped in the pycnocline layer and is transported to the right (e.g. (1.25, 0.7) m).  A large fraction of the energy however is transmitted through the layer.  Very little energy is reflected off the second layer, allowing the rest of the energy to reflect off the bottom of the domain.

The finite difference method is applied to the modified ocean density profile, and the beam-normalized percent difference of the energy flux magnitude is presented in figure~\ref{fig:fig8}(c).  The largest errors occur near the more abrupt transition layer.  The maximum percent difference in this region 28.1$\%$.  There is no consistent trend with regards to under or over estimating the energy flux.  Outside of the immediate region of the sharper transition, the percent difference is generally within 5$\%$.  It is also important to note that the method is able to capture and accurately determine the energy flux in the reflected, transmitted, and trapped internal waves outside of the highly nonlinear first reflection region.

\section{Conclusions} \label{section-conclusions}

We have presented two methods for calculating for internal waves the  instantaneous energy flux field using only  density perturbation field data.   Both methods are applicable to nonlinear stratifications: the first method,   a Green's function method,   uses convenient analytic density stratification profiles, while the second,  a finite difference method,  applies to arbitrary stratification profiles.  

Using our Green's function method we obtained the instantaneous energy flux from density perturbations for two buoyancy frequency profiles:  one linear in $z$ and the other where $N(z)^2 \propto \tanh(z)$.  The difference between the Green's function method and our direct numerical simulations is less than 3\% outside of regions containing significant nonlinearity.  Despite the Green's function method being based on linear theory, it accurately predicts the energy flux in the transmitted, reflected, and second harmonic beams, which involve significant nonlinearities. 

With our  finite difference method we showed how to   capture the energy flux in an internal wave field containing nonlinear interactions, wave beam reflections, and second harmonic beams for any buoyancy frequency profile $N(z)$. This method was  compared with the Green's function method and direct numerical simulations, and again the errors are less than 3\% for most of the domain. 
 
The two methods presented here and in \citet{allshouse16} allow detailed studies of the entire instantaneous energy flux field for internal wave field data, as contrasted with methods that yield a single global conversion rate or a time-averaged result.  Our methods can be used to determine the instantaneous velocity perturbation, pressure perturbation, and energy flux fields from density perturbation data obtained in experiments using synthetic schlieren or light attenuation measurements.  We emphasize that the methods  require only the density perturbation field over time and the background buoyancy frequency profile. Application to ocean observations will be possible provided a time-varying density perturbation field can be measured.  The methods assume the flow is two dimensional, but future work could extend the method to weakly three-dimensional flows as in ocean applications.

The Matlab GUI $``EnergyFlux"$ developed in ~\citet{allshouse16} is extended to include the methods discussed in this paper.  The GUI requires density perturbation data, domain coordinates, time step size, and the $N(z)$ profile. A manual and tutorial that reproduces the results in this work is provided to make possible straightforward applications of the methods presented here.

\section*{Acknowledgement}
We thank Robert Moser for helpful discussions about the numerical simulations. The computations were done at the Texas Advanced Computing Center. MRA and HLS were supported by the Office of Naval Research MURI Grant N000141110701, while FML and PJM were supported by the U.S. Department of Energy, Office of Science, Office of Fusion Energy Sciences, under Award Number DE-FG02-04ER-54742.


\end{document}